%% file: main.tex
\documentclass[conference]{IEEEtran}
\IEEEoverridecommandlockouts
\usepackage{cite}
\usepackage{amsmath,amssymb,amsfonts}
\usepackage{algorithmic}
\usepackage{authblk}
\usepackage{graphicx}
\usepackage{textcomp}
\usepackage{xcolor}
\usepackage{subcaption}
\usepackage{algorithm}
\usepackage{makecell}
\usepackage{hyperref}
\usepackage[normalem]{ulem}
\def\BibTeX{{\rm B\kern-.05em{\sc i\kern-.025em b}\kern-.08em
    T\kern-.1667em\lower.7ex\hbox{E}\kern-.125emX}}

\begin{document}

\title{K-PACT: Kernel Planning for Adaptive Context Switching — A Framework for Clustering, Placement, and Prefetching in Spectrum Sensing\\}

\author[1]{H. Umut Suluhan\thanks{
This material is based on research sponsored Defense Advanced Research Projects Agency (DARPA) under contract number HR001123C0130.}}
\author[2]{Jiahao Lin}
\author[1]{Serhan Gener}
\author[3]{Chaitali Chakrabarti}
\author[2]{Umit Ogras}
\author[1]{Ali Akoglu}
\affil[1]{Electrical and Computer Engineering Department, The University of Arizona, Tucson, AZ, USA}
\affil[2]{Department of Electrical and Computer Engineering, University of Wisconsin–Madison, Madison, WI, USA}
\affil[3]{School of Electrical, Computer, and Energy Engineering, Arizona State University, Tempe, AZ, USA}
\affil[ ]{\textit{\{suluhan,gener,akoglu\}@arizona.edu, \{jlin445,uogras\}@wisc.edu, chaitali@asu.edu}}

\maketitle

\begin{abstract}
Efficient wideband spectrum sensing requires rapid evaluation and re-evaluation of signal presence and type across multiple subchannels. These tasks involve multiple hypothesis testing, where each hypothesis is implemented as a decision-tree workflow containing compute-intensive kernels, including FFT, matrix operations, and signal-specific analyses. Given the dynamic nature of the spectrum environment, ability to quickly switch between hypotheses is essential for maintaining low-latency, high-throughput operation.
This work assumes a coarse-grained reconfigurable architecture consisting of an array of processing elements (PEs), each equipped with a local instruction memory (IMEM) capable of storing and executing kernels used in spectrum sensing applications. We propose a planner tool that efficiently maps hypothesis workflows onto this architecture to enable fast runtime context switching with minimal overhead.
The planner performs two key tasks: clustering temporally non-overlapping kernels to share IMEM resources within a PE sub-array, and placing these clusters onto hardware to ensure efficient scheduling and data movement. By preloading kernels that are not simultaneously active into same IMEM, our tool enables low-latency reconfiguration without runtime conflicts. It models the planning process as a multi-objective optimization, balancing trade-offs among context switch overhead, scheduling latency, and dataflow efficiency.
We evaluate the proposed tool in simulated spectrum sensing scenario with 48 concurrent subchannels. Results show that our approach reduces off-chip binary fetches by 207.81$\times$, lowers average switching time by 98.24$\times$, and improves per-subband execution time by 132.92$\times$ over baseline without preloading. These improvements demonstrate that intelligent planning is critical for adapting to fast-changing spectrum environments in next-generation radio frequency systems.

\end{abstract}



\input{tex/1_introduction}

\input{tex/2_background}
\input{tex/3_algorithm}
\input{tex/4_experimental_setup}
\input{tex/5_results}

\input{tex/6_related_work}

\input{tex/7_conclusion}

\section*{Acknowledgment}

This material is based on research sponsored Defense Advanced Research Projects Agency (DARPA) under contract number HR001123C0130. The U.S. Government is authorized to reproduce and distribute reprints for Governmental purposes, notwithstanding any copyright notation thereon. The views and conclusions contained herein are those of the authors and should not be interpreted as necessarily representing the official policies or endorsements, either expressed or implied, of the Defense Advanced Research Projects Agency (DARPA) or the U.S. Government.

Dr. Akoglu and Dr. Ogras have disclosed an outside interest in DASH Tech IC to the University of Arizona and University of Wisconsin, respectively.  Conflicts of interest resulting from this interest are being managed by the respective universities in accordance with their policies.

\bibliographystyle{IEEEtran}
\bibliography{refs.bib}

\end{document}

%% file: tex/1_introduction.tex
\section{Introduction}



Spectrum sensing is a fundamental process in cognitive radio networks that identifies the presence and spectrum usage within a specific geographical area~\cite{yucek2009survey}. It can detect unauthorized use and facilitate dynamic spectrum access by opportunistically allocating unused frequency bands to secondary users without disrupting primary user communications. Spectrum sensing involves analyzing multiple dimensions of the spectrum space, time, frequency, and location to ensure efficient and non-intrusive spectrum utilization.



In advanced spectrum sensing scenarios, the objective is multiple hypothesis testing, where the goal is to distinguish among several possible states of spectrum occupancy. 
A hypothesis is a decision tree based workflow composed of a series of tests that involve computationally intensive kernels such as FFT, matrix inversion, and matrix multiplication used for spectrum analysis based on components such as energy detection (ED), orthogonal frequency-division multiplexing (OFDM) analysis, and cyclic prefix (CP) detection~\cite{axell2012spectrum}. These components are essential for accurately identifying various signal types and characteristics across multiple subchannels.
Instead of just deciding whether a user is present or absent, multiple hypothesis testing allows for finer-grained classification of different signal types across numerous subchannels.
As a result, the detection problem evolves into a comprehensive test with a significantly larger hypothesis space, increasing computational complexity exponentially with the number of subchannels~\cite{axell2012spectrum,zeng2010review}.

Software-defined or autonomous radio frequency (RF) systems must continuously assess signal relevance and dynamically switch context on the datapath, replacing low-priority data and tasks with new hypotheses and their associated workflows~\cite{moy2015software,ulversoy2010software}.
Therefore, spectrum sensing requires a hardware system capable of executing numerous decision trees to support evaluating multiple signals across subchannels using multiple hypotheses. The hardware should not only support concurrent processing of multiple subchannels but also enable seamless reconfiguration to incorporate new hypotheses as signals arrive dynamically. \textit{Such dynamic execution requires the ability to configure the hardware and continuously reconfigure for new decision tree scenarios in real-time for wideband spectrum sensing}. 

A coarse-grained array of processing elements (PEs), where each PE supports a subset of the spectrum sensing kernels, enables setting up pipelines to test multiple hypotheses concurrently~\cite{chen2025canalis,podobas2020survey}. 
This concurrency can be achieved by sharing the PE-level instruction memories (IMEMs) among multiple co-located kernels and rapidly context switching among them.
This facilitates rapid reconfiguration on the array as the hypotheses change by allowing the system to load kernel binaries into the local IMEMs of each PE.
A runtime planning software tool is needed to bring the context into the hardware ahead of time so that real-time reconfiguration can be realized by switching between kernels that share the IMEMs of the sub-array.
The proposed tool involves clustering and placement phases. Clustering identifies the group of kernels that should share IMEMs, while the placement determines the physical location of the clusters in the PE array to enable concurrent execution of multiple hypotheses.
The planning software solves a multi-objective optimization problem that balances the trade-off between context switching, scheduling latency, and data flow latency with its clustering and placement decisions.
This paper proposes K-PACT\footnote{The source code can be found in \href{https://github.com/UA-RCL/K-PACT}{https://github.com/UA-RCL/K-PACT}}, a software planning tool that (1) clusters temporally independent kernels within shared PEs, and then (2) assigns them to distinct IMEM banks. Since these kernels are not active at the same time, K-PACT enables rapid context switching within each PE without any conflicts. Leveraging the proposed planning tool, we simulate a spectrum sensing scenario with 48 concurrent subchannels and perform tradeoff studies on key performance metrics, including average switching time, scheduling time, and execution time per subband.
The proposed software planning tool reduces off-chip kernel binary fetches by 207.81$\times$, resulting in a 98.24$\times$ decrease in overall switching time compared to the baseline. By loading kernels ahead of time, the tool minimizes the need for costly runtime scheduling decisions. Instead, a dynamic resource management scheme is used to identify appropriate locations for incoming decision trees, which reduces scheduling time by 73.77$\times$. Collectively, these optimizations yield a 132.92$\times$ reduction in subband execution time, demonstrating that intelligent software planning, coupled with IMEM sharing across PEs, is critical in rapidly changing spectrum environments.

%% file: tex/2_background.tex
\section{Background}

\begin{figure}[t]
     \centering
         \centering
         \includegraphics[width=0.7\linewidth]{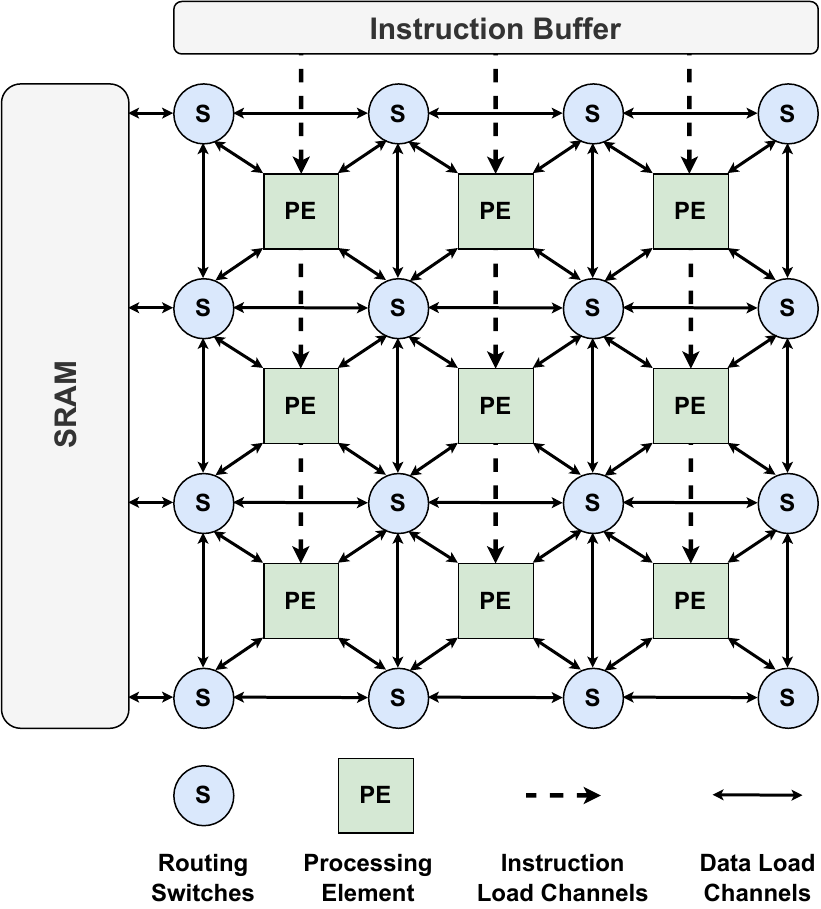}
        \caption{Reference systolic array architecture for K-PACT}
        \vspace{-8pt}
        \label{fig:systolic-array}
\end{figure}

\subsection{Hardware Architecture}

We utilize a modified version of the recently introduced systolic array known as FluxSPU~\cite{chen2025canalis} illustrated in Figure~\ref{fig:systolic-array}, 
as \textit{a representative PE-array architecture}. FluxSPU is a spatial architecture where PEs are arranged in a two-dimensional array and connected via bi-directional switches. In the original design, multiple PEs can be grouped to form a processing kernel capable of executing compute-intensive tasks. Depending on the task, each PE is preloaded with a dedicated instruction binary generated in advance. In the modified implementation used in this work, each PE is equipped with multiple IMEM banks, which allows the storage of several kernel binaries within a single PE. This design allows rapid switching between kernels at runtime by dynamically selecting the active IMEM bank through a global manager. We define three types of task-switching: no, soft, and hard switch, with varying degrees of reconfiguration cost.

\subsubsection{No Switch}

In a \emph{no switch} scenario, the required kernel is already actively loaded in the IMEM of the PEs, and ready to execute without any reconfiguration. As a result, no task-switching operation is required. The kernel begins execution immediately, introducing minimal overhead since it avoids off-chip memory access latency and switches between different IMEM banks.

\subsubsection{Soft Switch}

When a new kernel is initiated for a set of PEs, and the binary for the new kernel is already present in the IMEMs of the PEs, the runtime software identifies its location and begins execution from that point. In this scenario, the PEs execute the new set of instructions directly from IMEM without the need to reload binaries from off-chip memory, resulting in minimal switching overhead. This type of task transition is referred to as a \emph{soft switch}.


\subsubsection{Hard Switch}
\label{sub:hard-switch}
When the binary for the new kernel is not present in the IMEM of the PEs on the array, it must be fetched from off-chip memory and loaded into the on-chip Instruction Buffer and then into the PE array. If a cluster of PEs with sufficient available IMEM is found, the switching process involves searching for these PEs and loading the new kernel binary into their IMEMs. Suppose no such cluster is available. In that case, the system must verify that no suitable configuration exists, allocate a new cluster of PEs, and load the kernel binary into their IMEMs.
In cases where the array lacks enough free space to allocate a new group, one or more existing kernel binaries must be evicted from the IMEMs to make room. This process introduces additional overhead due to eviction, reallocation, and binary loading from off-chip memory, resulting in the highest switching cost. Compared to a soft switch, a \emph{hard switch} incurs more substantial overhead due to the need for memory transfers and potential reconfiguration. However, intelligent PE allocation and IMEM management strategies can mitigate this overhead by optimizing resource usage and reducing the kernel relocation frequency.

\begin{figure}[t]
    \centering
    \includegraphics[width=0.95\linewidth]{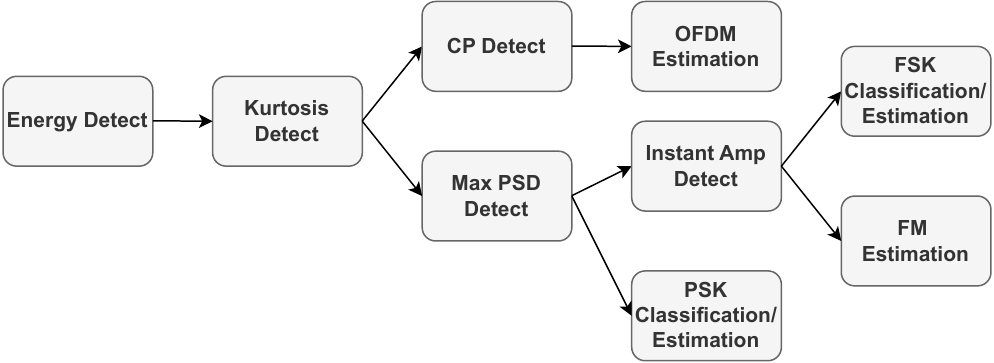}
    \caption{Example decision tree structure}
    \vspace{-8pt}
    \label{fig:decision-tree}
\end{figure}

\subsection{Spectrum Sensing Scenarios and Deployment}
\label{sub:spectrum}
We represent a hypothesis over a spectrum bandwidth as a decision tree, as shown in Figure~\ref{fig:decision-tree}. In this structure, each node corresponds to a processing kernel, and edges capture the conditional dependencies between kernels. Signals arrive dynamically in a frame-based fashion, and each is processed through a corresponding decision tree, referred to as a subband. A collection of such trees constitutes a processing scenario, where multiple signal channels are handled in parallel. In our experimental setup, the system processes up to 48 subbands concurrently at any given time. The execution flow is determined by threshold-based decisions at each node, the outcome of a kernel determines whether the signal is dropped or which subsequent kernel is invoked. The example in Figure~\ref{fig:decision-tree} begins with the \textit{Energy Detect} kernel. If the signal's energy exceeds a predefined threshold, the \textit{Kurtosis Detect} kernel is executed. Based on its output, the signal is either dropped or passed to the \textit{CP Detect} or \textit{Max PSD Detect} kernel. This conditional execution continues until the signal is either discarded or successfully classified or estimated. 

%% file: tex/3_algorithm.tex
\section{K-PACT Algorithm}


\begin{algorithm}[t]
\caption{Kernel clustering based on temporal independence} \label{alg:clustering}
\begin{algorithmic}[1]
    \STATE \textbf{Inputs: $kernels, imem\_sizes$}
    \STATE \textbf{Outputs: $clusters$}
    \STATE $i \gets 0$
    \STATE $clusters \gets []$
    \WHILE{not all kernels are clustered}
        \STATE $clusters[i] \gets []$ 
        \STATE $score \gets []$ 
        \FOR{each kernel $k_1$ in kernels}
            \STATE $score[k_1] \gets 0$ 
            \FOR{each kernel $k_2$ in kernels, where $k_1 \neq k_2$}
                \STATE Compute temporal independence of $k_1$ and $k_2$
                \STATE Update $score[k_1]$ based on independence measure
            \ENDFOR
        \ENDFOR
        \STATE Sort $score$ in descending order
        
        \STATE $s\_k \gets score[0]$ 
        \STATE $j \gets 0$
        \WHILE{$j <$ number of kernels}
            \STATE $c\_k \gets kernels[j]$ 
            \IF{$s\_k \neq c\_k$}
                \IF{$s\_k$ and $c\_k$ mutually exclusive}
                    \STATE Absorb $c\_k$ into $clusters[i]$
                    \STATE Remove $c\_k$ from kernels
                \ENDIF
            \ENDIF
            \STATE $j \gets j + 1$
        \ENDWHILE
        \STATE $i \gets i + 1$
    \ENDWHILE
        \FOR{each cluster $c$ in clusters}
            \STATE $imem\_size \gets 0$ 
            \FOR{each kernel $k$ in $c$}
                \STATE $imem\_size \gets imem\_size + imem\_sizes[k]$
            \ENDFOR
            \STATE $new\_cluster \gets []$
            \STATE $i \gets 0$
            \WHILE{$imem\_size > imem\_size\_limit$}
                \STATE Remove tail kernel $k$ from $c$ 
                \STATE $new\_cluster[i] \gets k$
                \STATE $i \gets i + 1$
            \ENDWHILE
            \STATE Append $new\_cluster$ to $clusters$
        \ENDFOR
\end{algorithmic}
\end{algorithm}

After analyzing system behavior across different environments, each reflecting distinct hypotheses and subband conditions, we can collect sample records representing kernel activity in response to these environmental changes. Each sample contains information such as the start ($s_i$) and end ($e_i$) times of a kernel $\mathcal{K}_i$. Using this information, the next step is to cluster temporally independent kernels where their binaries can be stored in the same IMEM, enabling rapid switching in future executions under similar environmental conditions.

\subsection{Problem Statement}
The core objective is to group $N$ kernels into $M$ clusters such that each cluster satisfies a user-defined IMEM size constraint while minimizing the total number of clusters $M$. Each cluster may contain up to $N'$ temporally independent kernels, where $1 \leq N' \leq N$. Kernels that are active concurrently must be duplicated across different clusters to ensure conflict-free execution within the PE-array. The cluster formation (Algorithm~\ref{alg:clustering}) can be formalized as in Equation~\ref{eq:cluster}, where $\mathcal{K}_i$ denotes the current kernel under consideration, and $\mathcal(s_i, e_i)$ is the start and end time for the kernel. $\mathcal{C}_j$ denotes the current cluster that kernels are grouped into. $\text{I}_{\text{req}}$ denotes the IMEM requirement imposed by the kernel and $\text{I}_{\text{lim}}$ is the user-defined upper bound for IMEM size.
Formally, a kernel $\mathcal{K}_i$ can be assigned to a cluster $\mathcal{C}_j$ if its start time $s_i$ is after the end time $e_{i-1}$ of the previous kernel $\mathcal{K}_{i-1}$, and the total IMEM requirement ($\text{I}_{\text{req}}$) of all kernels in $\mathcal{C}_j$ remains below the user-defined limit $\text{I}_{\text{lim}}$.

\begin{equation}
\label{eq:cluster}
\begin{aligned}
&\mathcal{K}_i \in \mathcal{C}_j \iff \left( e_{i-1} < s_i \right) \land \sum_{\mathcal{K}_k \in \mathcal{C}_j} \mathrm{I}_{\mathrm{req}}(\mathcal{K}_k) < \mathrm{I}_{\mathrm{lim}}, \\
&\forall \, \mathcal{K}_i^{(s_i, e_i)} \in \{\mathcal{K}_1, \dots, \mathcal{K}_N\}, \quad \forall \, \mathcal{C}_j \in \{\mathcal{C}_1, \dots, \mathcal{C}_M\}
\end{aligned}
\end{equation}

This formulation enables efficient reuse of clusters and supports fast reconfiguration. While the proposed method is demonstrated using spectrum sensing as a case study, it can be generalized to any application domain where kernel executions are temporally independent.





\subsection{Clustering Phase}



The clustering process referring to Algorithm~\ref{alg:clustering} starts with assessing the temporal non-overlap between all kernel pairs within the designated kernel set to derive a ranking metric (lines 5-14). The kernel exhibiting the highest degree of temporal non-overlap is designated as the seed kernel, forming the foundation of the initial cluster (lines 15-16). 
Then, a greedy aggregation strategy iteratively incorporates new kernels that demonstrate temporal independence with the seed kernel into the same cluster (lines 17-27).
This initial clustering phase operates under the unbounded IMEM capacity assumption. If the kernel under consideration is not suitable to fit into any of the existing clusters due to temporal overlap with the resident kernels in them, a new cluster is created (lines 5-6).
The ranking process repeats, and the clustering continues until all kernels are assigned to clusters (lines 5-29). A subsequent clipping operation is performed to meet user-specified IMEM capacity constraints (lines 30-43). 
If the IMEM utilization of a cluster surpasses the limit, kernels are iteratively removed and reassigned to new clusters until all clusters satisfy the IMEM capacity requirements (lines 37-41). 

\begin{figure}[t]
    \centering
    \includegraphics[width=0.5\linewidth]{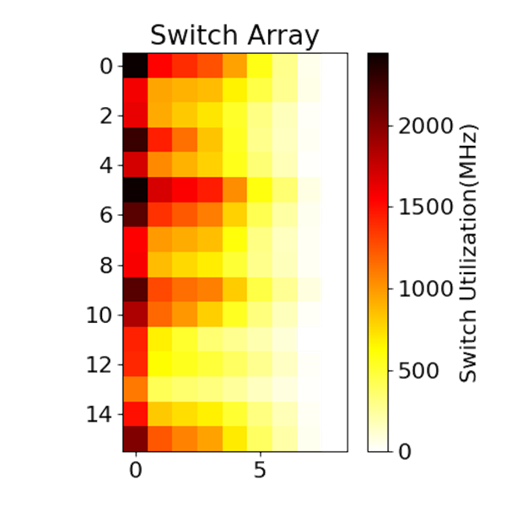}
    \caption{Deployment for a simple scenario showing the stress on the first column of switches}
    \label{fig:switch-congestion}
    \vspace{-8pt}
\end{figure}

\subsection{Placement Phase}
After temporal-independence-driven clustering, the proposed tool places the clusters onto the PE array to minimize the time to feed data to the kernel instances while maximizing the number of concurrently executing decision trees. 
The PE array receives input data from SRAM buffers, and data flows horizontally to downstream kernels through routing switches. When multiple decision trees execute concurrently, dataflow aware placement becomes essential. 
Placing frequently accessed entry kernels of decision trees far from the SRAM buffers increases the propagation times and causes routing switch congestion as illustrated with Figure~\ref{fig:switch-congestion}. Positioning the entry blocks close to the SRAM and distributing the remaining kernels across the subregions of the PE array is crucial to mitigate this overhead. Following this strategy, the planning software forms specialized clusters for frequently accessed kernels and places them near the SRAM buffers. The remaining kernels are grouped separately and distributed across the rest of the PE array. This placement scheme helps alleviate overall congestion, as high-frequency kernels can access input data with minimal latency.


\subsection{Design Decisions}
\subsubsection{Reducing Hard Switches}
As the system receives new decision trees at runtime, the dynamic placer parses the decision tree from head to tail nodes as summarized in Section~\ref{sub:spectrum}. If the currently visited node (kernel) is not available on the PE array, then the dynamic placer determines if this kernel can be absorbed into one of the existing clusters that satisfy the clustering and IMEM size constraints. Otherwise, it generates a new cluster on the PE array or applies the eviction rules as described in Section~\ref{sub:hard-switch} to load a kernel binary into the PE array, considering the hard-switch latency penalty. On a cold system start, all distinct kernels visited in a decision tree will observe a hard-switch penalty. To minimize hard-switch intensive programming of the kernels on the PE array, one option is to pre-initialize the clusters on the PE array based on known and commonly used spectrum sensing decision tree deployment scenarios. We evaluate two forms of pre-initialization called \textit{Pre-initialized Placement (PIP)} and \textit{Fixed Pre-initialized Placement (FPIP)}.

\begin{figure}[t]
    \centering
    \includegraphics[width=0.6\linewidth]{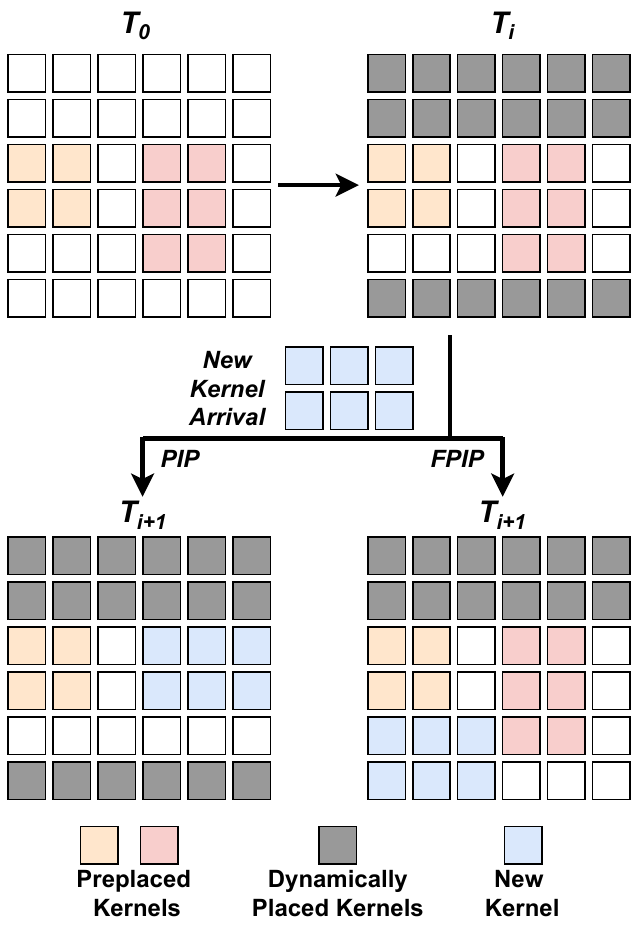}
    \caption{Pre-placement followed by dynamic placement options}
    \label{fig:ds3-exec}
    \vspace{-8pt}
\end{figure}

Suppose we pre-initialize the PE array with two kernels as denoted by orange and red regions in Figure~\ref{fig:ds3-exec} at time $T_{0}$. Assume that the dynamic placement gradually loads the PE array with a new set of kernels, and the state of the PE array at time $T_{i}$ involves the two pre-initialized kernels along with dynamically placed kernels labeled as gray shaded regions. At this instance, suppose that the PE array is out of spatial space and IMEM lines are fully occupied. In this case, an eviction event occurs when a new kernel that is not already loaded in the array arrives. For \textit{PIP} case, we allow the dynamic resource manager to evict the preplaced kernels for accommodating the newly arriving kernel. While this approach still leverages the advantages of \textit{PIP}, allowing dynamic resource manager to evict these preplaced kernels diminishes the benefits of profiling-guided pre-placement. In contrast, \textit{Fixed Pre-initialized Placement (FPIP)} prevents the dynamic placer from evicting kernels placed initially by the software planning tool.


\subsubsection{Choosing IMEM Size}

A larger IMEM size allows grouping more number of kernels within a single cluster, enabling more frequent soft switches instead of hard ones. However, increasing the IMEM size results in a larger area per PE. Beyond a certain threshold, increasing the IMEM size leads to a significant increase in the total PE-array area despite reducing the number of clusters. As a result, an optimal IMEM size must be determined to efficiently accommodate the given set of trees while minimizing the total area of the PE-array. 

%% file: tex/4_experimental_setup.tex
\section{Experimental Setup}


We use an open-source domain-specific system-on-chip simulator (DS3)~\cite{arda2020ds3} to simulate a set of scenarios for evaluation. DS3 models the hardware architecture shown in Figure~\ref{fig:systolic-array} and a dynamic scheduler \textit{(DP)}. The dynamic scheduler continuously monitors and manages kernels programmed across the PE array. If an instance of a newly arrived kernel is not present, it generates real-time decisions for kernel placement, routing, and resource allocations. By default, the simulation begins with a cold start with an empty array, allowing \textit{DP} to place kernels as they arrive.

\noindent\textbf{Scheduling modes:} 
We implement \textit{Baseline} version of \textit{DP} that cannot perform any soft switches, and relies only on hard switches. \textit{DP} version has the ability to form clusters and execute soft switches. We implement the \textit{PIP} and \textit{FPIP} pre-initialization strategies and refer to them as \textit{PIP+DP} and \textit{FPIP+DP}, respectively. In the \textit{PIP+DP} approach, \textit{DP} can evict kernels from the preplaced regions, whereas \textit{FPIP+DP} strictly preserves those placements. 

\noindent\textbf{Performance metrics:} 
We collect performance metrics to demonstrate the benefit of pre-initialized placement and compare different integration approaches. For every kernel switch event, we record whether it is a hard/soft switch or no switch and refer to them as \texttt{Hard Switch Count}, \texttt{Soft Switch Count}, and \texttt{No Switch Count}, respectively. To showcase the effect of switching behavior for the workload execution, we measure \texttt{Average Instruction Load Time}, \texttt{Average Data Load Time}, and \texttt{Average Switching Time}. \texttt{Average Instruction Load Time} refers to the average time taken to load kernel binaries into IMEM banks and calculated as following:
\begin{equation}
\begin{split}
\frac{
N_{\text{Hard}} \cdot O_{\text{Hard}} + 
N_{\text{Soft}} \cdot O_{\text{Soft}} + 
N_{\text{No}} \cdot O_{\text{No}}
}{
N_{\text{Hard}} + N_{\text{Soft}} + N_{\text{No}}
}
\end{split}
\end{equation}
Where $N_{\text{Hard}}$, $N_{\text{Soft}}$, and $N_{\text{No}}$ denotes the number of hard, soft, and no switches, respectively. $O_{\text{Hard}}$, $O_{\text{Soft}}$, and $O_{\text{No}}$ are the overheads associated with each switch type. \texttt{Average Data Load Time} records data loading event for each kernel, and averages with the total number of switch count. 

Additionally, we measure scheduling time overhead by the dynamic scheduler referred to as \texttt{Average Scheduling Time}. It encapsulates the time taken by dynamic scheduler to check whether kernel is preloaded or not and perform placement decisions in case of a hard switch. To quantify the effect of switching and scheduling time on overall performance, we provide \texttt{Average Execution Time Per Subband} metric, which is calculated by dividing total makespan by the number of processed subbands.


\begin{figure}[t]
    \centering
    \includegraphics[width=0.95\linewidth]{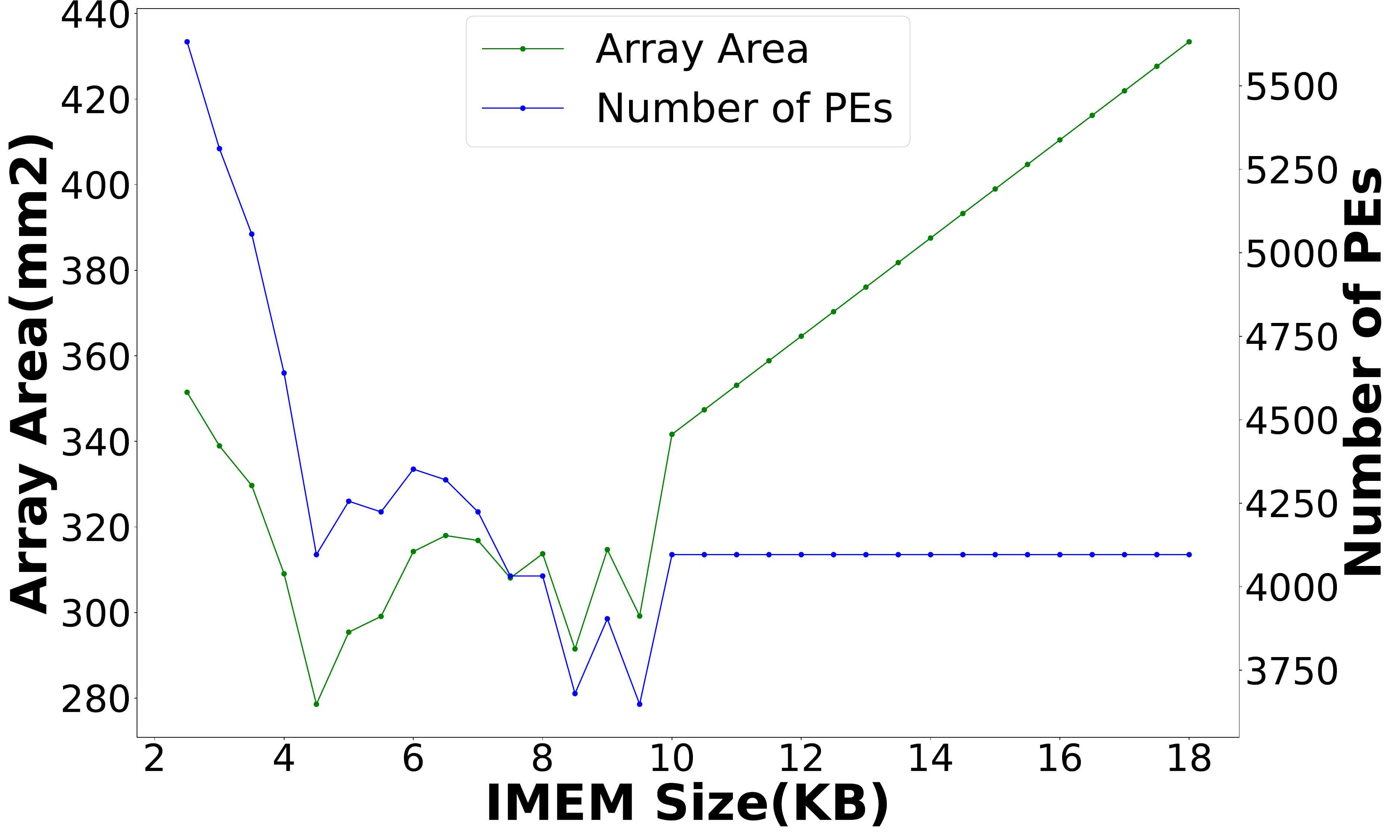}
    \caption{IMEM size and array area tradeoff}
    \vspace{-8pt}
    \label{fig:2d-plot}
\end{figure}

\noindent\textbf{IMEM Size:} 
An optimal IMEM size must be determined to efficiently accommodate the given set of trees while minimizing the total array area. 
To this end, we sweep the IMEM size and perform clustering and placement repeatedly. Figure~\ref{fig:2d-plot} illustrates this analysis, highlighting the trade-off between IMEM size, the number of PEs, and total array area. In this study, the number of PEs and total area are computed following each clustering and placement iteration. The total area is calculated by multiplying the number of PEs and the area of a single PE, then adding the area occupied by the SRAM buffers, as shown in Equation~\ref{eq:total_area}. PE and SRAM buffer areas are estimated post-synthesis using GlobalFoundries' 12nm process. 
\begin{equation}
\label{eq:total_area}
\text{Total Area} = N_{\text{PE}} \times (A_{\text{logic}} + A_{\text{IMEM}}) + N_{\text{rows}} \times A_{\text{SRAM}}
\end{equation} 
Where $N_{\text{PE}}$ denotes the number of PEs, $A_{\text{logic}}$ represents the fixed logic area. $A_{\text{IMEM}}$ is a function of IMEM size, and gradually increases as the IMEM size grows. $N_{\text{rows}}$ denotes the number of rows of the PE-array, and $A_{\text{SRAM}}$ is the single buffer area. Increasing the number of rows helps reduce congestion but results in higher area overhead. For the given example scenario, $4.5KB$ IMEM achieves the optimal balance, minimizing the total array area while accommodating all required kernels. Increasing the IMEM size further leads to significant area overhead due to the increased per-PE area, outweighing the benefits of clustering. Therefore, we fix the IMEM size to 4.5KB and preload clusters to pre-determined locations in the PE-array in the rest of the paper.

%% file: tex/5_results.tex
\section{Experimental Results}

\begin{figure*}[ht]
     \centering
  \setlength\tabcolsep{1pt}
  \begin{tabular}{cc}
     \begin{subfigure}[c]{0.5\textwidth}
         \centering
         \includegraphics[width=\textwidth]{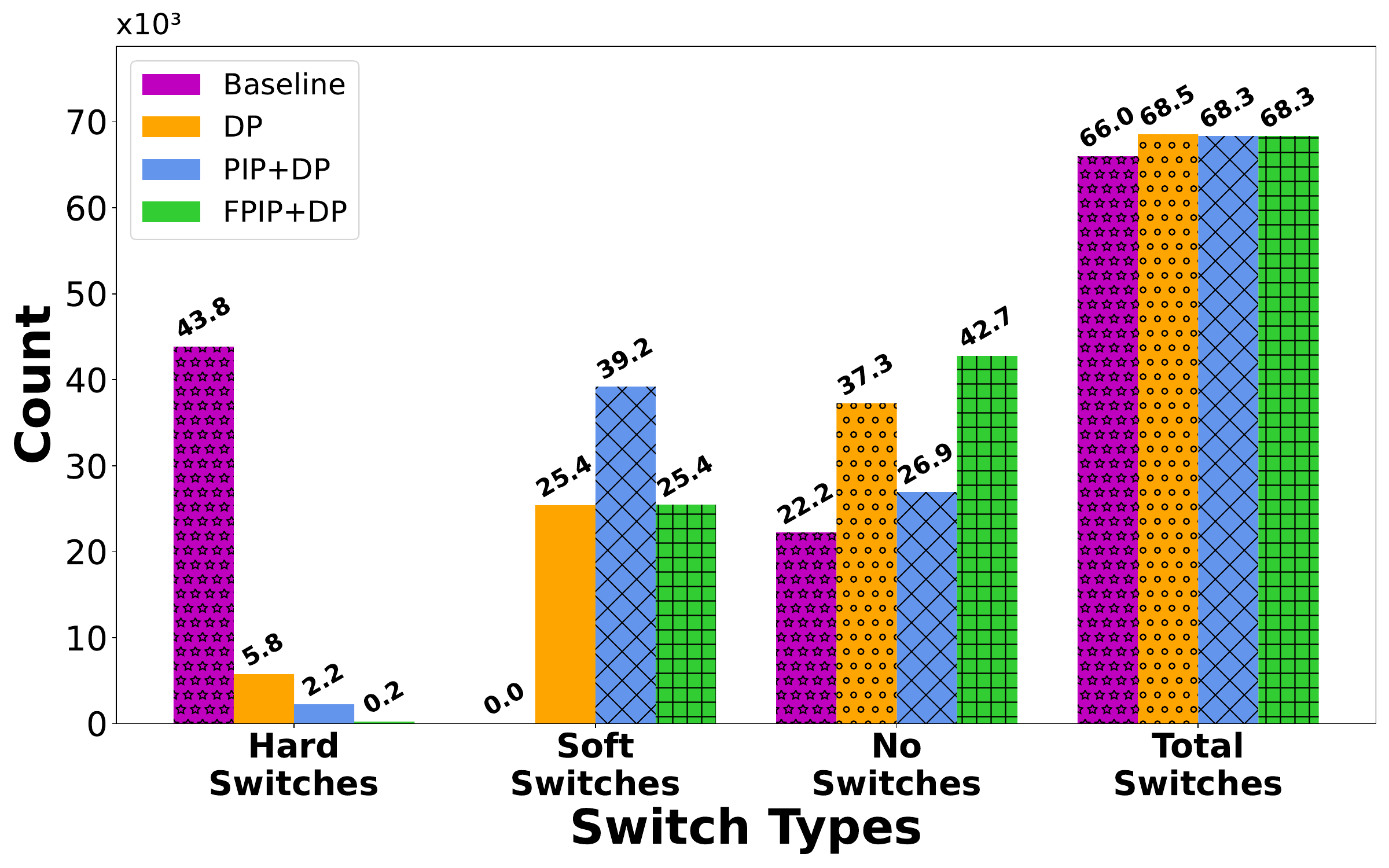}
        \caption{Switch count analysis}
         \label{fig:switch-count}
     \end{subfigure}&
     \begin{subfigure}[c]{0.5\textwidth}
         \centering
         \includegraphics[width=\textwidth]{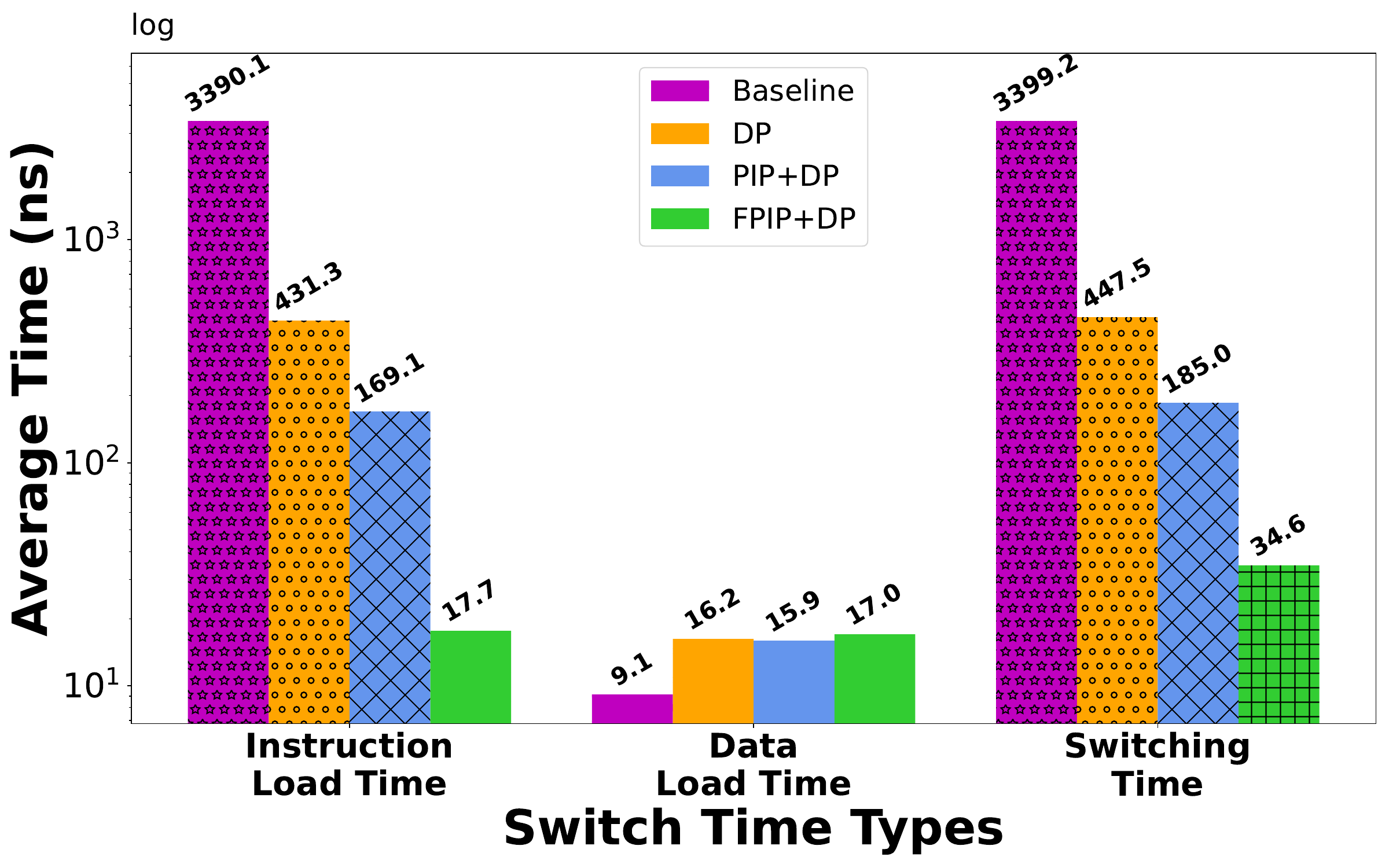}
        \caption{Switch time breakdown and analysis}
         \label{fig:switch-time}
         
     \end{subfigure}
     \end{tabular}
        \caption{Switching analysis based on 48 subbands deployment scenario.}
        \label{fig:switch-count-time-analysis}
        \vspace{-8pt}
\end{figure*}

\subsection{Benefit of Pre-Initialized Placement}

Figure~\ref{fig:switch-count} presents an empirical analysis of the four aforementioned kernel placement strategies and illustrates the trade-off between hard and soft switches. 
Analyzing hard switch trends shows that the \textit{Baseline}, which lacks rapid switching capability, results in the highest number of hard switches. By enabling soft switch functionality, the system achieves a $7.61\times$ reduction in hard switches due to effective IMEM sharing across PEs. Further, transitioning from \textit{DP} to \textit{PIP+DP}, where kernels are preloaded in advance, leads to an additional $2.58\times$ decrease in hard switches. However, \textit{PIP+DP} allows \textit{DP} to evict the preloaded kernels. Avoiding eviction for the pre-loaded kernels with \textit{FPIP+DP} results in a $10.56\times$ reduction compared with the \textit{PIP+DP}. Overall, the combination of soft switching and intelligent kernel planning delivers a $207.81$× improvement in hard switch count compared to the \textit{Baseline}.

The reduction in hard switch count leads to a more frequent use of soft and no switches, as the required kernels are already programmed into the IMEM banks of PEs. When we examine the soft and no switch counts for \textit{DP}, \textit{PIP+DP}, and \textit{FPIP+DP}, we find that \textit{PIP+DP} exhibits the highest number of soft switches while requiring the fewest no switches. This behavior is primarily driven by the hard switches that occur in \textit{PIP+DP}. When a preloaded kernel is evicted, it triggers a chain of hard switch events because the required kernels are no longer available in the array. Throughout this chain of hard switches, the \textit{PIP+DP} strategy adapts to the system's evolving needs and loads the necessary kernels into the array. This process results in dynamic runtime clustering and IMEM sharing across the array, allowing \textit{PIP+DP} to increasingly rely on soft switches for kernel transitions. In contrast, \textit{FPIP+DP} avoids evicting any preloaded kernels and makes extensive use of no switches, as the required kernels are already placed and ready for execution.
Overall, the \textit{FPIP+DP} approach fully exploits rapid switching capability by preloading the necessary kernels ahead of time. Although \textit{PIP+DP} achieves a reduction in hard switch count compared to \textit{DP} and \textit{Baseline}, evicting preloaded kernels results in additional hard switches, ultimately limiting its overall benefit.

Figure~\ref{fig:switch-time} illustrates the overall time spent during switches for the same experiment. The y-axis represents average time in nanoseconds, while the x-axis depicts three different switch time metrics, instruction load time, data load time, and average switch time. Analyzing the instruction load time highlights that reducing hard switches significantly lowers the overhead caused by off-chip binary fetches. With \textit{PIP+DP} and \textit{FPIP+DP} modes enabled, the system effectively utilizes rapid context switching through soft and no switch operations. As a result, instruction load time improves by $20.04\times$ and $2.55\times$ under \textit{PIP+DP}, compared to \textit{Baseline} and \textit{DP}, respectively. Further gains are achieved with \textit{FPIP+DP}, which restricts the eviction of the preloaded kernels and retains frequently used kernels in the array. This approach leads to $191.53\times$ and $24.36\times$ reduction in instruction load time relative to \textit{Baseline} and \textit{DP}, respectively. Pre-initialized placement strategies effectively exploits the hardware’s switching capabilities by bringing contexts in ahead of time, thereby minimizing costly off-chip binary fetch.

We observe that the \textit{Baseline} approach incurs the least overhead among all configurations on the data load time. This result stems from its reliance on frequent and expensive hard switches, which significantly delay kernel execution due to loading binaries from off-chip storage. As a result, the array remains underutilized, and routing channels handle fewer active kernels. With reduced contention, data is delivered faster to each kernel. When examining the relationship between \textit{DP} and \textit{PIP+DP}, we observe a lower data load time with \textit{PIP+DP}, primarily because it sustains execution using higher soft switches and fewer number of no switches. This results in less routing switch congestion and faster data load time in \textit{PIP+DP} compared to \textit{DP}. On the other hand, \textit{FPIP+DP} shows the highest data load time. Since most required kernels are already programmed, execution proceeds through no switches, leading to heavy array utilization and increased pressure on data routing paths.

When analyzing overall switching time, we find that instruction load time is the dominant contributor for the \textit{Baseline}, \textit{DP}, and \textit{PIP+DP} approaches. \textit{PIP+DP} improves the overall switching time by $18.37\times$ and $2.41\times$ compared to \textit{Baseline} and \textit{DP}, respectively. The improvement with \textit{FPIP+DP} is more substantial, where $98.24\times$ and $12.93\times$ speedup is observed against \textit{Baseline} and \textit{DP} strategies. Frequent hard switches bottlenecks system utilization by introducing delays from off-chip binary fetches. In contrast, with \textit{FPIP+DP}, the bottlenecks shift toward data load time. This shift highlights the need for routing-aware software planning tools that can smartly manage routing channel usage to reduce contention, especially under high array utilization.

\begin{figure}[t]
    \centering
    \includegraphics[width=0.95\linewidth]{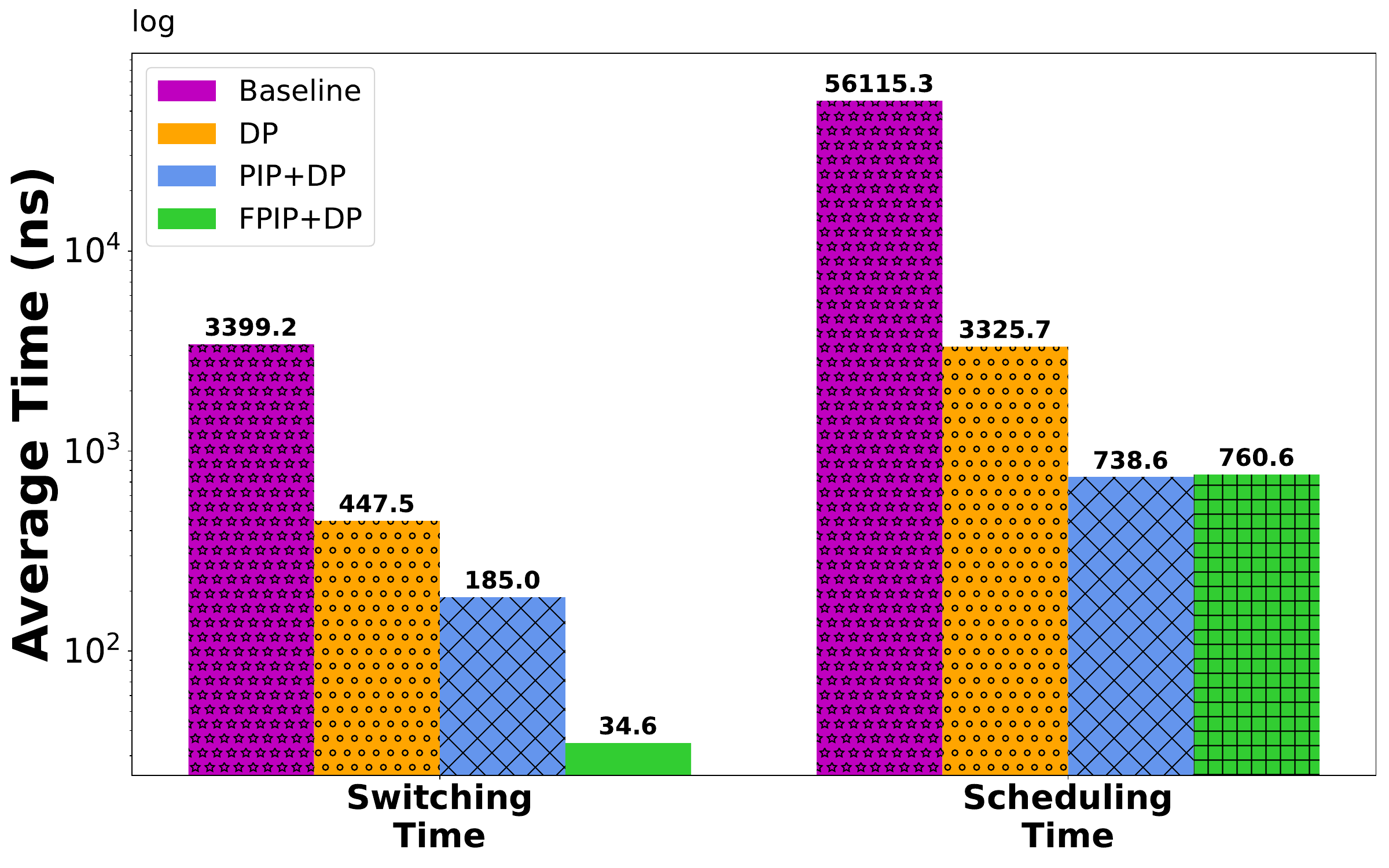}
    \caption{Scheduling time analysis}
    \vspace{-8pt}
    \label{fig:sched-time}
\end{figure}

When we analyze the scheduling overhead presented in Figure~\ref{fig:sched-time}, we observe that the \textit{Baseline} approach incurs the highest overhead due to the excessive number of hard switches. Each hard switch triggers the scheduler to locate a suitable placement for the incoming kernel, which involves a costly PE availability check across all potential array locations. As the number of hard switches decreases with the introduction of \textit{DP} and \textit{PIP+DP}, we observe $16.87\times$ and $75.97\times$ shorter scheduling time, respectively. An interesting trend arises when comparing \textit{PIP+DP} and \textit{FPIP+DP}. Although \textit{FPIP+DP} performs fewer hard switches and is expected to achieve lower scheduling overhead, its eviction cost is significantly higher due to its inability to evict any preloaded kernels. This limitation can be effectively mitigated by marking the coordinates of preloaded kernels as unavailable during initialization. Another trend is that scheduling time dominates the switching time for all execution approaches. This is mainly due to eviction events, where the arriving kernel cannot be placed because of limited space or unavailable IMEM lines. In such scenarios, the scheduler must scan all clusters to identify a candidate for eviction, which introduces a significant delay.

\begin{table}[t]
    \centering
    \begin{tabular}{|c|c|c|c|}
        \hline
        \makecell{Execution\\Approach} & 
        \makecell{Execution\\Time (us)} & 
        \makecell{Speedup wrt\\\textit{Baseline}} & 
        \makecell{Speedup wrt\\\textit{DP}} \\
        \hline
        Baseline & 431.622 & 1x & N/A\\
        DP & 9.943 & 43.40x & 1x\\
        PIP+DP & 3.865 & 111.67x & 2.57x\\
        FPIP+DP & 3.247 & 132.92x & 3.06x\\
        \hline
    \end{tabular}
    \caption{Average execution time per subband}
    \label{tab:exec-time}
\end{table}

To evaluate the combined impact of switching and scheduling time on overall system performance, we present the average execution time per subband in Table~\ref{tab:exec-time}. The \textit{Baseline} approach results in the longest execution time due to frequent hard switches and binary fetches from off-chip storage. Introducing soft switch capabilities yields a significant $43.40\times$ improvement, as kernels can now be initiated via soft switches. With the integration of pre-initialized placement in \textit{PIP+DP} and \textit{FPIP+DP}, execution time is further reduced, achieving $111.67\times$ and $132.92\times$ speedups over the \textit{Baseline}, respectively. This gain arises from the ability to have context readily available ahead of time, allowing most operations to proceed using soft or no switches.
When comparing speedups relative to \textit{DP}, \textit{FPIP+DP} achieves only a $1.19\times$ improvement over \textit{PIP+DP}, despite delivering a $4.17\times$ reduction in switching time. This limited overall gain is due to increased scheduling overhead caused by kernel eviction events. Given that scheduling time dominates switching time (as shown in Figure~\ref{fig:sched-time}), its contribution to total execution time is more significant. Nevertheless, the execution time still benefits from the reduction in switching time and yields better execution times. This outcome highlights that in systems with rapid switching capabilities, efficient scheduling is essential. Minimizing costly eviction events is critical for maintaining performance in array-based architectures.

%% file: tex/6_related_work.tex
\section{Related Work}
A commonly used approach for managing a wideband spectrum channel is to divide it into multiple subchannels as smaller frequency segments within the broader spectrum. Some of these subchannels may be active, while others remain unoccupied. The goal of spectrum sensing is to determine the occupancy status of each subchannel and classify the type of signals present~\cite{axell2012spectrum}. As signal activity changes over time, the system must be capable of adapting rapidly by replacing or modifying hypotheses (e.g., decision trees) to reflect new spectrum conditions. This dynamic model of execution demands both hardware and software support for real-time context switching at nanosecond-scale granularity, particularly for processing GHz-wide bands~\cite{konatham2020real, xia2022design, muzaffar2024review, eappen2020survey, ivanov2021probabilistic}.
FPGAs and CGRAs are well-suited to such dynamic workloads due to their inherent reconfigurability and parallelism. However, existing CAD and compiler flows for these architectures are typically optimized for static, single-context application execution. In the case of FPGAs, clustering algorithms are designed to group LUT-level operations into Configurable Logic Blocks (CLBs) to minimize routing congestion and improve timing. These approaches aim to localize connections and minimize inter-CLB routing based on a single application's dependency graph, and are guided by metrics such as wirelength~\cite{gort2012analytical, kim2011simpl, vercruyce2017liquid}, routability~\cite{chen2017ripplefpga, abuowaimer2018gplace3}, and timing~\cite{li2018utplacef, chen2020clock, zhu2022high}. Some methods target multi-objective optimization for better trade-offs among these goals~\cite{li2019elfplace, elgammal2021rlplace}.
Similarly, CGRA mapping techniques also focus on optimizing a single application execution context. These tools exploit the computational capabilities of PEs through intelligent dataflow graph (DFG) partitioning and pipelining. Approaches range from heuristic-based mapping~\cite{zhao2017static, dave2018ramp, canesche2020traversal} to those based on Reinforcement Learning~\cite{liu2018data, zhuang2022towards, kong2023mapzero}, genetic algorithms~\cite{kojima2020genmap}, and hierarchical schemes~\cite{wijerathne2021himap}. In all these cases, the assumption is that a single application or kernel remains mapped and active over a fixed execution period.
In contrast, K-PACT is designed to meet the requirements of highly dynamic, multi-context workloads in real-time spectrum sensing. Instead of clustering operations for a single static context, K-PACT focuses on maximizing PE reuse across temporally non-overlapping kernels from multiple decision trees. By analyzing temporal independence among kernels, K-PACT clusters them to share PEs and associated instruction memory, thereby enabling nanosecond-scale context switching without the need for full hardware reconfiguration.
This shift in clustering objective from spatial packing for a single application (as in traditional FPGA/CGRA mapping) to temporal co-location for rapid context switching across multiple applications is critical for applications that demand real-time adaptability. K-PACT’s unique clustering and mapping strategy bridges the gap between the need for real-time responsiveness and efficient hardware utilization, setting it apart from existing FPGA and CGRA compilation approaches.

%% file: tex/7_conclusion.tex
\section{Conclusion}

Spectrum sensing is a vital management scheme allowing secondary users to use under-utilized spectrum bands. However, it is a challenging task requiring high throughput and instantaneous detection capability to handle highly dynamic spectrum environments. These requirements necessitate a processor with fast switching capability and powerful software abstractions for mapping the kernels to computational resources across the hardware efficiently. 
In this work, we introduce K-PACT, a software planning tool that assigns temporally independent kernels to shared PEs with distinct IMEM banks, enabling rapid and conflict-free context switching by ensuring the kernels do not execute concurrently.
We simulate a heavy-workload spectrum sensing scenario consisting of 48 concurrent subbands with an event-based simulator for studying the benefits of the proposed software planning tool.
Simulation results show that K-PACT reduces the instruction load time by 191.53$\times$ and the overall execution time of a subband by 132.92$\times$ through leveraging fast-switching capability of the hardware. This improvement can be extended to application domains other than spectrum sensing, essentially addressing the requirements of multiple-context and rapidly changing workloads. 
As future work, we plan to develop routing-aware clustering and placement algorithms alongside with efficient scheduling heuristics to further improve the performance gains through fast context switching.

%% file: main.bbl
\begin{thebibliography}{10}
\providecommand{\url}[1]{#1}
\csname url@samestyle\endcsname
\providecommand{\newblock}{\relax}
\providecommand{\bibinfo}[2]{#2}
\providecommand{\BIBentrySTDinterwordspacing}{\spaceskip=0pt\relax}
\providecommand{\BIBentryALTinterwordstretchfactor}{4}
\providecommand{\BIBentryALTinterwordspacing}{\spaceskip=\fontdimen2\font plus
\BIBentryALTinterwordstretchfactor\fontdimen3\font minus \fontdimen4\font\relax}
\providecommand{\BIBforeignlanguage}[2]{{%
\expandafter\ifx\csname l@#1\endcsname\relax
\typeout{** WARNING: IEEEtran.bst: No hyphenation pattern has been}%
\typeout{** loaded for the language `#1'. Using the pattern for}%
\typeout{** the default language instead.}%
\else
\language=\csname l@#1\endcsname
\fi
#2}}
\providecommand{\BIBdecl}{\relax}
\BIBdecl

\bibitem{yucek2009survey}
T.~Yucek and H.~Arslan, ``A survey of spectrum sensing algorithms for cognitive radio applications,'' \emph{IEEE communications surveys \& tutorials}, vol.~11, no.~1, pp. 116--130, 2009.

\bibitem{axell2012spectrum}
E.~Axell, G.~Leus, E.~G. Larsson, and H.~V. Poor, ``Spectrum sensing for cognitive radio: State-of-the-art and recent advances,'' \emph{IEEE signal processing magazine}, vol.~29, no.~3, pp. 101--116, 2012.

\bibitem{zeng2010review}
Y.~Zeng, Y.-C. Liang, A.~T. Hoang, and R.~Zhang, ``A review on spectrum sensing for cognitive radio: challenges and solutions,'' \emph{EURASIP journal on advances in signal processing}, vol. 2010, pp. 1--15, 2010.

\bibitem{moy2015software}
C.~Moy and J.~Palicot, ``Software radio: a catalyst for wireless innovation,'' \emph{IEEE Communications Magazine}, vol.~53, no.~9, pp. 24--30, 2015.

\bibitem{ulversoy2010software}
T.~Ulversoy, ``Software defined radio: Challenges and opportunities,'' \emph{IEEE Communications Surveys \& Tutorials}, vol.~12, no.~4, pp. 531--550, 2010.

\bibitem{chen2025canalis}
K.-Y. Chen, T.~Mason~Nelson, A.~Khadem, M.~Fayazi, S.~S.~V. Singapuram, R.~Dreslinski, N.~Talati, H.-S. Kim, and D.~Blaauw, ``Canalis: A throughput-optimized framework for real-time stream processing of wireless communication,'' \emph{ACM Transactions on Reconfigurable Technology and Systems}, vol.~17, no.~4, pp. 1--32, 2025.

\bibitem{podobas2020survey}
A.~Podobas, K.~Sano, and S.~Matsuoka, ``A survey on coarse-grained reconfigurable architectures from a performance perspective,'' \emph{IEEE Access}, vol.~8, pp. 146\,719--146\,743, 2020.

\bibitem{arda2020ds3}
S.~E. Arda, A.~Krishnakumar, A.~A. Goksoy, N.~Kumbhare, J.~Mack, A.~L. Sartor, A.~Akoglu, R.~Marculescu, and U.~Y. Ogras, ``Ds3: A system-level domain-specific system-on-chip simulation framework,'' \emph{IEEE Transactions on Computers}, vol.~69, no.~8, pp. 1248--1262, 2020.

\bibitem{konatham2020real}
S.~R. Konatham, R.~Maram, L.~Romero~Cort{\'e}s, J.~H. Chang, L.~Rusch, S.~LaRochelle, H.~Guillet~de Chatellus, and J.~Aza{\~n}a, ``Real-time gap-free dynamic waveform spectral analysis with nanosecond resolutions through analog signal processing,'' \emph{Nature communications}, vol.~11, no.~1, p. 3309, 2020.

\bibitem{xia2022design}
M.~Xia, N.~Zhao, G.~Yin, R.~Yang, X.~Chen, C.~Feng, and M.~Dong, ``A design of real-time data acquisition and processing system for nanosecond ultraviolet-visible absorption spectrum detection,'' \emph{Chemosensors}, vol.~10, no.~7, p. 282, 2022.

\bibitem{muzaffar2024review}
M.~U. Muzaffar and R.~Sharqi, ``A review of spectrum sensing in modern cognitive radio networks,'' \emph{Telecommunication Systems}, vol.~85, no.~2, pp. 347--363, 2024.

\bibitem{eappen2020survey}
G.~Eappen and T.~Shankar, ``A survey on soft computing techniques for spectrum sensing in a cognitive radio network,'' \emph{SN Computer Science}, vol.~1, no.~6, p. 352, 2020.

\bibitem{ivanov2021probabilistic}
A.~Ivanov, K.~Tonchev, V.~Poulkov, and A.~Manolova, ``Probabilistic spectrum sensing based on feature detection for 6g cognitive radio: A survey,'' \emph{IEEE Access}, vol.~9, pp. 116\,994--117\,026, 2021.

\bibitem{gort2012analytical}
M.~Gort and J.~H. Anderson, ``Analytical placement for heterogeneous fpgas,'' in \emph{22nd international conference on field programmable logic and applications (FPL)}.\hskip 1em plus 0.5em minus 0.4em\relax IEEE, 2012, pp. 143--150.

\bibitem{kim2011simpl}
M.-C. Kim, D.-J. Lee, and I.~L. Markov, ``Simpl: An effective placement algorithm,'' \emph{IEEE Transactions on Computer-Aided Design of Integrated Circuits and Systems}, vol.~31, no.~1, pp. 50--60, 2011.

\bibitem{vercruyce2017liquid}
D.~Vercruyce, E.~Vansteenkiste, and D.~Stroobandt, ``Liquid: High quality scalable placement for large heterogeneous fpgas,'' in \emph{2017 International Conference on Field Programmable Technology (ICFPT)}.\hskip 1em plus 0.5em minus 0.4em\relax IEEE, 2017, pp. 17--24.

\bibitem{chen2017ripplefpga}
G.~Chen, C.-W. Pui, W.-K. Chow, K.-C. Lam, J.~Kuang, E.~F. Young, and B.~Yu, ``Ripplefpga: Routability-driven simultaneous packing and placement for modern fpgas,'' \emph{IEEE Transactions on Computer-Aided Design of Integrated Circuits and Systems}, vol.~37, no.~10, pp. 2022--2035, 2017.

\bibitem{abuowaimer2018gplace3}
Z.~Abuowaimer, D.~Maarouf, T.~Martin, J.~Foxcroft, G.~Gr{\'e}wal, S.~Areibi, and A.~Vannelli, ``Gplace3. 0: Routability-driven analytic placer for ultrascale fpga architectures,'' \emph{ACM Transactions on Design Automation of Electronic Systems (TODAES)}, vol.~23, no.~5, pp. 1--33, 2018.

\bibitem{li2018utplacef}
W.~Li, Y.~Lin, M.~Li, S.~Dhar, and D.~Z. Pan, ``Utplacef 2.0: A high-performance clock-aware fpga placement engine,'' \emph{ACM Transactions on Design Automation of Electronic Systems (TODAES)}, vol.~23, no.~4, pp. 1--23, 2018.

\bibitem{chen2020clock}
J.~Chen, Z.~Lin, Y.-C. Kuo, C.-C. Huang, Y.-W. Chang, S.-C. Chen, C.-H. Chiang, and S.-Y. Kuo, ``Clock-aware placement for large-scale heterogeneous fpgas,'' \emph{IEEE Transactions on Computer-Aided Design of Integrated Circuits and Systems}, vol.~39, no.~12, pp. 5042--5055, 2020.

\bibitem{zhu2022high}
Z.~Zhu, Y.~Mei, Z.~Li, J.~Lin, J.~Chen, J.~Yang, and Y.-W. Chang, ``High-performance placement for large-scale heterogeneous fpgas with clock constraints,'' in \emph{Proceedings of the 59th ACM/IEEE Design Automation Conference}, 2022, pp. 643--648.

\bibitem{li2019elfplace}
W.~Li, Y.~Lin, and D.~Z. Pan, ``elfplace: Electrostatics-based placement for large-scale heterogeneous fpgas,'' in \emph{2019 IEEE/ACM International Conference on Computer-Aided Design (ICCAD)}.\hskip 1em plus 0.5em minus 0.4em\relax IEEE, 2019, pp. 1--8.

\bibitem{elgammal2021rlplace}
M.~A. Elgammal, K.~E. Murray, and V.~Betz, ``Rlplace: Using reinforcement learning and smart perturbations to optimize fpga placement,'' \emph{IEEE Transactions on Computer-Aided Design of Integrated Circuits and Systems}, vol.~41, no.~8, pp. 2532--2545, 2021.

\bibitem{zhao2017static}
Z.~Zhao, W.~Sheng, W.~He, Z.~Mao, and Z.~Li, ``A static-placement, dynamic-issue framework for cgra loop accelerator,'' in \emph{Design, Automation \& Test in Europe Conference \& Exhibition (DATE), 2017}.\hskip 1em plus 0.5em minus 0.4em\relax IEEE, 2017, pp. 1348--1353.

\bibitem{dave2018ramp}
S.~Dave, M.~Balasubramanian, and A.~Shrivastava, ``Ramp: Resource-aware mapping for cgras,'' in \emph{Proceedings of the 55th Annual Design Automation Conference}, 2018, pp. 1--6.

\bibitem{canesche2020traversal}
M.~Canesche, M.~Menezes, W.~Carvalho, F.~S. Torres, P.~Jamieson, J.~A. Nacif, and R.~Ferreira, ``Traversal: A fast and adaptive graph-based placement and routing for cgras,'' \emph{IEEE Transactions on Computer-Aided Design of Integrated Circuits and Systems}, vol.~40, no.~8, pp. 1600--1612, 2020.

\bibitem{liu2018data}
D.~Liu, S.~Yin, G.~Luo, J.~Shang, L.~Liu, S.~Wei, Y.~Feng, and S.~Zhou, ``Data-flow graph mapping optimization for cgra with deep reinforcement learning,'' \emph{IEEE Transactions on Computer-Aided Design of Integrated Circuits and Systems}, vol.~38, no.~12, pp. 2271--2283, 2018.

\bibitem{zhuang2022towards}
Y.~Zhuang, Z.~Zhang, and D.~Liu, ``Towards high-quality cgra mapping with graph neural networks and reinforcement learning,'' in \emph{Proceedings of the 41st IEEE/ACM International Conference on Computer-Aided Design}, 2022, pp. 1--9.

\bibitem{kong2023mapzero}
X.~Kong, Y.~Huang, J.~Zhu, X.~Man, Y.~Liu, C.~Feng, P.~Gou, M.~Tang, S.~Wei, and L.~Liu, ``Mapzero: Mapping for coarse-grained reconfigurable architectures with reinforcement learning and monte-carlo tree search,'' in \emph{Proceedings of the 50th Annual International Symposium on Computer Architecture}, 2023, pp. 1--14.

\bibitem{kojima2020genmap}
T.~Kojima, N.~A.~V. Doan, and H.~Amano, ``Genmap: A genetic algorithmic approach for optimizing spatial mapping of coarse-grained reconfigurable architectures,'' \emph{IEEE Transactions on Very Large Scale Integration (VLSI) Systems}, vol.~28, no.~11, pp. 2383--2396, 2020.

\bibitem{wijerathne2021himap}
D.~Wijerathne, Z.~Li, A.~Pathania, T.~Mitra, and L.~Thiele, ``Himap: Fast and scalable high-quality mapping on cgra via hierarchical abstraction. in 2021 design, automation \& test in europe conference \& exhibition (date),'' 2021.

\end{thebibliography}
